\documentclass[aps,pre,showkeys,showpacs,groupedaddress,amsmath,amssymb,floatfix,twocolumn]{revtex4-1}

\usepackage{color} 
\usepackage{url}
\usepackage{todonotes}

\usepackage{lmodern}						

\begin{document}


\title{Pattern Recognition Approach to Violin Shapes of MIMO database}
\author{Thomas Peron$^{1}$}
\email{thomaskaue@gmail.com}
\author{Francisco A. Rodrigues$^{1}$}
\author{Luciano da F. Costa$^{2}$}
\affiliation{$^{1}$Institute of Mathematics and Computer Science, University of S\~ao Paulo, S\~ao Carlos, SP, Brazil}
\affiliation{$^{2}$S\~ao Carlos Institute of Physics, University of S\~ao Paulo, S\~ao Carlos, SP, Brazil}


\begin{abstract}
Since the landmarks established by the Cremonese school in the 16th century, the history of violin design has been marked by experimentation. While great effort has been invested since the early 19th century by the scientific community on researching violin acoustics, substantially less attention has been given to the statistical characterization of how
the violin shape evolved over time. In this paper we study the morphology of violins retrieved from the Musical Instrument Museums Online (MIMO) database -- the largest freely accessible platform 
providing information about instruments held in public museums. From the violin images, 
we derive a set of measurements that reflect relevant geometrical features
of the instruments. The application of Principal Component Analysis (PCA) uncovered similarities between 
violin makers and their respective copyists, as well as among luthiers belonging to the same family lineage, in the context of historical narrative. Combined with a time-windowed approach, thin plate splines visualizations revealed that the average violin outline has remained mostly stable over time, not adhering to any particular trends of design across different periods in music history. 
\end{abstract}

\maketitle 

\section{Introduction}

Music is among the most abstract of the arts, which can be a reason 
for its widespread appreciation.  As such, it becomes specially 
difficult to identify, compare and discuss its content, properties, 
quality, and relationships.  Compare it with, for instance, painting.  
In this case, it becomes possible to devise objective approaches to 
develop all the aforementioned analyses
(e.g. how similar to a Brussels griffon is the dog in van Eyck 
Arnolfini's portrait?).
Used for music production, musical instruments have varying 
properties such as resonances, tessiture, harmonic content, and so 
on.  Even though the final appreciation of a given instrument, as well 
as its playability, depends on subjective aspects of human perception 
and cognition, much can be learnt about instruments by applying 
the scientific method to them, which involves measuring, analysing 
and modeling their physical properties~\cite{woodhouse2014acoustics}.  

One of the most known classical instruments, which often appears 
in most genres of music, is the violin.  It seems to have been derived from
the bowed instruments from the equestrian cultures in Asia and 
undergone a long history of changes and adaptations, reaching its
modern form in Europe, particularly northern Italy, especially
with Stradivarius and Guarneri~\cite{boyden1990history,hutchins1983history,bachmann2013encyclopedia}.  
In a violin, the sound 
energy is produced by the friction between the bow and the strings, then 
reaching the resonance box through the bridge, where it is transferred to 
the surrounding air by the plate vibration.  Several elements contribute 
to the sound properties and quality of a violin, including the employed 
materials, the bow and strings, and the three-dimensional shape and dimensions 
of the body and arm~\cite{woodhouse2014acoustics}. In particular, the resonances are strongly related to 
the geometric Chladni patterns that are formed in the plate and body of the violin for 
different fundamental frequencies~\cite{hutchins1983history,gough2007violin}.

The history of violin research encompasses many disciplines related to different 
aspects of the instrument, such as vibration analysis~\cite{hutchins1983history,woodhouse2014acoustics}, material science~\cite{nagyvary2006wood,nagyvary2009mineral,tai2017chemical}, and psychoacoustics~\cite{fritz2014soloist,fritz2017listener}. One of the earliest scientists to perform experiments with the violin was Felix Savart~\cite{hutchins1983history}. Among his many contributions are the first studies on the vibration modes, including the first reports 
on applying the method by Chladni in order to reveal the nodal and anti-nodal regions in 
violin plates for different resonance frequencies. Many other prominent physicists were 
also involved with the research of bowed string instruments. For instance, it was Helmholtz 
who first uncovered the characteristic wave form produced by a string under the action 
of a bow~\cite{hutchins1983history,woodhouse2014acoustics}. His work strongly influenced Lord Rayleigh, who made several breakthroughs on vibrational analysis and laid the foundation for modern 
experimental and theoretical acoustics~\cite{hutchins1983history}. In the early 20th century, Raman carried out
remarkably detailed experiments with violins and cello strings~\cite{hutchins1983history,raman1920experiments}. Not only the Indian 
scientist shed light on the interplay between many parameters associated to violin playability (distance of the bow to the bridge, bow speed, bowing force, etc.), but also 
elucidated some aspects of the so-called ``wolf-note''~\cite{hutchins1983history,raman1916wolf}. For additional information on the history of violin research we refer the reader to~\cite{hutchins1983history,woodhouse2014acoustics,katz2006violin} and references therein.
 

Probably one of the reasons for the great interest, including the scientific community,
on the violin is the fascination surrounding famous luthiers like Stradivari and Guarneri, 
who were able to craft remarkable instruments. However, despite all the efforts by the 
mentioned scientists and others in trying to understand the physical properties of the violin, one aspect has received relatively little attention along time: the shape of the instrument's body. As remarked  by Chitwood~\cite{chitwood2014imitation}, the violin shape has been
intentionally ignored in modern acoustic research, in part due to experimental practicity. 
Indeed, many of the early experiments with bowed strings were performed with the strings 
isolated from the instruments. 

\begin{figure}[!tpb]
\begin{center}
\includegraphics[width=1.0\linewidth]{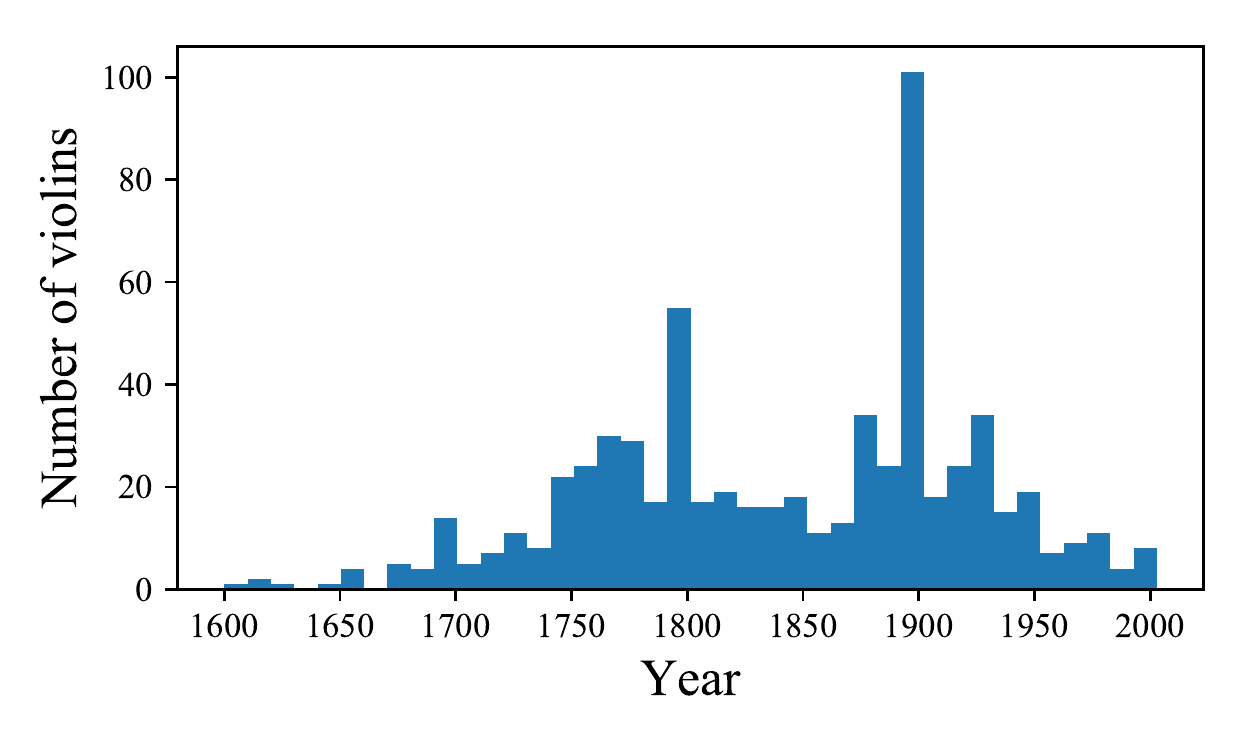}
\end{center}
\caption{Number of violins over time in the database used in this paper.}
\label{fig:histogram}
\end{figure}

These simplifications regarding the violin shape constrained more accurate modeling 
of physical properties of violins.  Indeed, it remains an unanswered question
how much shape differences can impact respective acoustical properties. Another 
question that naturally arises in this context is how distinguishable 
instruments fabricated by different luthiers can be. In other words, 
given a set of violins, is it possible to infer the authorship 
based solely on features extracted from the body outline? Furthermore, is it possible to 
infer, based on shape, how a given luthier or school influenced 
other instrument makers? Recently, a step forward in answering these questions 
was taken by Chitwood~\cite{chitwood2014imitation}. This author carried out a study which, 
to the best of our knowledge, is the first to systematically analyze a large
dataset of violin shapes. By employing pattern recognition techniques, it was found 
that the design of violins indeed correlates with the historical time line. Hierarchical 
clustering also revealed that famous luthiers tend to group in the same clusters
with their respective copyists~\cite{chitwood2014imitation}. 

In this paper, we extend some concepts in the methodology of~\cite{chitwood2014imitation} in order to morphologically analyze the violin shapes retrieved from another database, namely the
Musical Instrument Museums Online (MIMO)~\cite{MIMO}. The interest in this database
resides in the fact that it consists in the largest freely accessible 
platform that provides information about instruments held in public museums. Therefore, the
availability of the data and the significance of the collections provide motivation for a thorough statistical characterization and systematic comparison of the violins held in the museums associated to the MIMO platform. In order to accomplish this task, we define
a set of measures that reflect important geometrical features of the violin instrument. After defining these morphological indicators, 
the distribution of violins in the attribute space is obtained via Principal Component Analysis (PCA), whereby a significant relationship between luthiers and their respective copyists is revealed. Moreover, low statistical variability is found among violins fabricated by luthiers belonging 
to the same family lineage. Interesting results are also reported concerning the country of fabrication. In particular, great dispersion in the PCA space 
is observed for violins by French makers, whereas instruments fabricated in other countries tend exhibit higher levels of clustering.  Time-series analysis of the MIMO database were
also carried out in order to characterize the evolution of the geometrical features of violins
across different periods in classical music. In particular, by employing a time-windowed approach along with a thin plate splines method~\cite{costa2001shape}, we show that the average
violin shape has remained predominantly stable over time, in agreement with results
derived with PCA.








This paper is organized as follows: In Sec.~\ref{sec:materials_and_methods} we describe
in detail the MIMO database and the criteria used to filter the violin images
used in our study. The features to be extracted from the images are also defined in this section. Sec.~\ref{sec:discussion} is devoted to the discussion on the statistical 
similarity of violin parts. Sec.~\ref{sec:PCA} shows the results concerning 
PCA, followed by Sec.~\ref{sec:temporal_analysis_morphing}, which contemplates the 
analysis of the temporal evolution of violin measures as well as the time-evolving 
morphing via thin-plate splines technique. Finally, our conclusions are presented
in Sec.~\ref{sec:conclusions}.

\section{Materials and Methods}
\label{sec:materials_and_methods}

This section first describes in detail of 
the content displayed in the MIMO database as well as 
the criteria used to filter the images. Subsequently, 
we present the methodology proposed for the image processing 
and features extraction.

\subsection{The MIMO database and information filtering}

The Musical Instrument Museums Online (MIMO)~\cite{MIMO} is a platform that 
provides free and easy access to collections of musical instruments in important European museums. The information provided for each instrument comprises, for instance, year of fabrication, place of production, maker, museum, and images of the
instrument. In this paper, we analyze images retrieved by using
``violin'' as keyword in the MIMO's website. The respectively obtained instruments are filtered so that only images of non-rotated instruments are selected. A total of 726 violins out of 1300 instruments was chosen for this work, spanning the interval between 1600-2003. 

Some violins in the database have some of the above mentioned fields missing.  Figure~\ref{fig:histogram} shows the distribution along time of the 658 violins incorporating year of fabriation. The remaining elements either have imprecise data, e.g ``violin fabricated in the 18th century'', or no year is provided at all.   Other information, such as instrument maker 
and place of production, are also missing sometimes. Tables~\ref{table:makers} and~\ref{table:countries} show the most common countries of fabrication and makers along with the respective number of elements with missing data.

\begin{table}
\begin{tabular}{cc}
\hline 
Instrument maker & Number of violins\tabularnewline
\hline 
\hline 
Carl Friedrich Hopf & 12\tabularnewline
\hline 
Otto Glass & 12\tabularnewline
\hline 
Jerome Thibouville & 9\tabularnewline
\hline 
Jean-Baptiste Vuillaume & 8\tabularnewline
\hline 
Carleen Hutchins & 7\tabularnewline
\hline 
Others & 426\tabularnewline
\hline 
Unknown & 252\tabularnewline
\hline 
\end{tabular}\caption{Table showing the most frequent violin makers in the MIMO database.}
\label{table:makers}

\end{table}

\begin{table}
\begin{tabular}{cc}
\hline 
Country & Number of violins\tabularnewline
\hline 
\hline 
France & 233\tabularnewline
\hline 
Germany & 232\tabularnewline
\hline 
Italy & 37\tabularnewline
\hline 
Belgium & 17\tabularnewline
\hline 
Czech Republic & 13\tabularnewline
\hline 
Others & 38\tabularnewline
\hline 
Unknown & 156\tabularnewline
\hline 
\end{tabular}\caption{Countries of fabrication of violins in the MIMO database.}
\label{table:countries}

\end{table}

\subsection{Violin Shape Measurements}

\begin{figure}[!tpb]
\begin{center}
\includegraphics[width=0.8\linewidth]{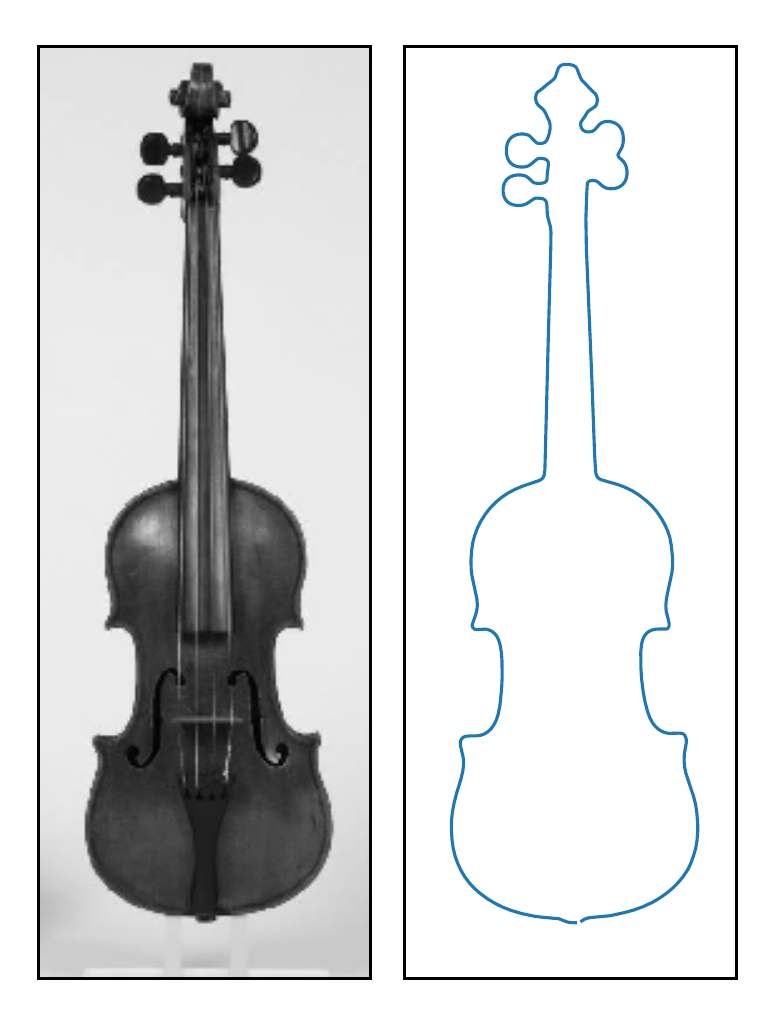}
\end{center}
\caption{Example of contour extraction. (left) Original image in gray-scale and (right) extracted contour. Violin image reprinted with permission by the Musikinstrumentenmuseums der Universit\"at Leipzig. }
\label{fig:contour_example}
\end{figure}

One of our goals is to comparatively analyze
physical properties of violins as estimated from their respective images (see Fig.~\ref{fig:contour_example} for an example of contour extraction).   To this end, we first need to define, given a violin contour, what measures need to be derived.  In particular, good measurements should be able to capture the most relevant geometrical features of the violins, allowing a more objective and precise analysis and comparison.  Natural and more direct/intuitive measurement candidates include the \emph{lengths} of parts of violins.  Moreover, curvature-based measures also provide a particularly~\cite{costa2001shape} powerful way to represent and characterize shapes: it is invariant to rotation and translation, it preserves most of shape information (is invertible up to a rigid-body transformation), can be used to detect corners, and is compatible with human intuition.  Curvature is, therefore, also considered in this work. We can thus organize the measurements adopted here into two major groups: (a) length-based and (b) curvature-based measures.

\subsubsection{Length-Based Measurements and control points determination}

\begin{figure}[!tpb]
\begin{center}
\includegraphics[width=1.0\linewidth]{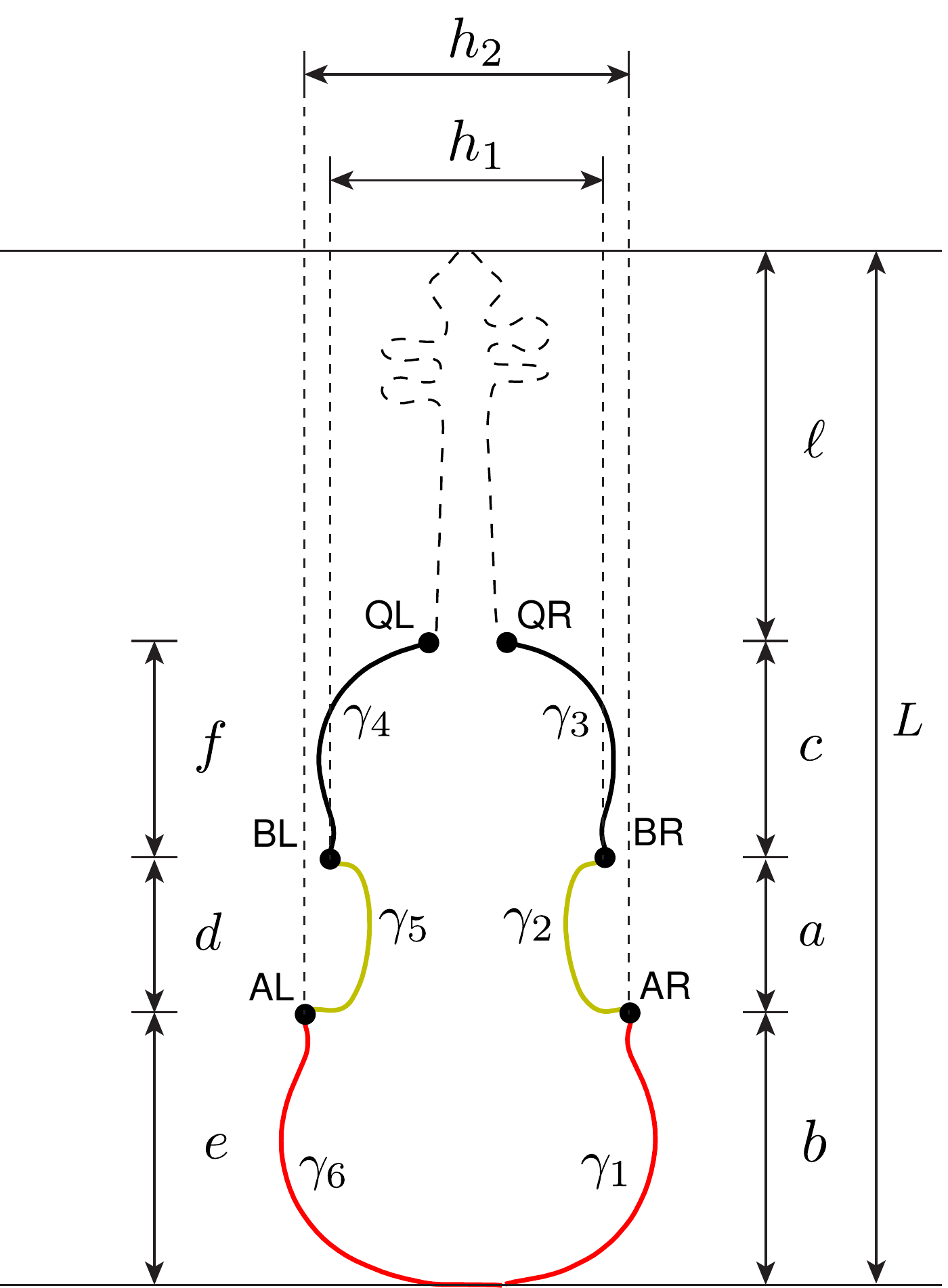}
\end{center}
\caption{Definition of the measures extracted from the violin contours. Dashed line in the violin contour represents the part which encompasses the 
fingerboard, pegbox, tuning pegs and scroll and which is not considered in the analysis. Symbols $\gamma_i$ ($i=1,...,6$) denote the contour segments defined by the control points.}
\label{fig:measures}
\end{figure}

A basic and natural way to describe a violin is by means of the length of its parts. Indeed, the length of an instrument can be a signature of a particular epoch or luthier~\cite{hutchins1983history,StradivariLiveNWork}. For instance, violins of the Barroque period are significantly shorter than ones from the 19th century, 
period at which the instruments were lengthened so as to improve the sound power output in larger 
concert halls~\cite{hutchins1983history}.  In order to capture such structural changes in the processing of the images, the first set of measurements we adopt is based on the length of different violin sections. Figure~\ref{fig:measures} illustrates in detail the length measurements to be used for violin characterization. 

One important first step in obtaining length-based measurements consists in defining reference points that can be used to organize the overal violin outline.  These points, henceforth called \emph{control points}, are shown in Fig~\ref{fig:measures} as AL, AR, BL, BR.
It turns out that the control points are points in the violin contour with significantly higher curvature value than the rest, as exemplified in Fig.~\ref{fig:explaining_detection}, where we show a typical contour extracted from a violin image. 

The values $a$ to $f$ quantify the extension of the upper, central and lower parts of the instrument; $h_1$ and $h_2$ stand for the width between respective control points; $\ell$ is the size of the region encompassing the neck, pegbox and the scroll; and $L$ corresponds to the total violin length.

\subsubsection{Curvature-Based Measurements}

Another relevant aspect that
differentiates one instrument from the other is the shape itself which is, in principle, a dimensionless property. Therefore, in order to characterize geometrical variations in violin contours, we also calculate the average curvature values $\bar{s}_i$ (see Appendix~\ref{sec:appendix_curvature}) for each of the segments $\gamma_i$ defined in Fig.~\ref{fig:measures}. 

\begin{figure}[!tpb]
\begin{center}
\includegraphics[width=1.0\linewidth]{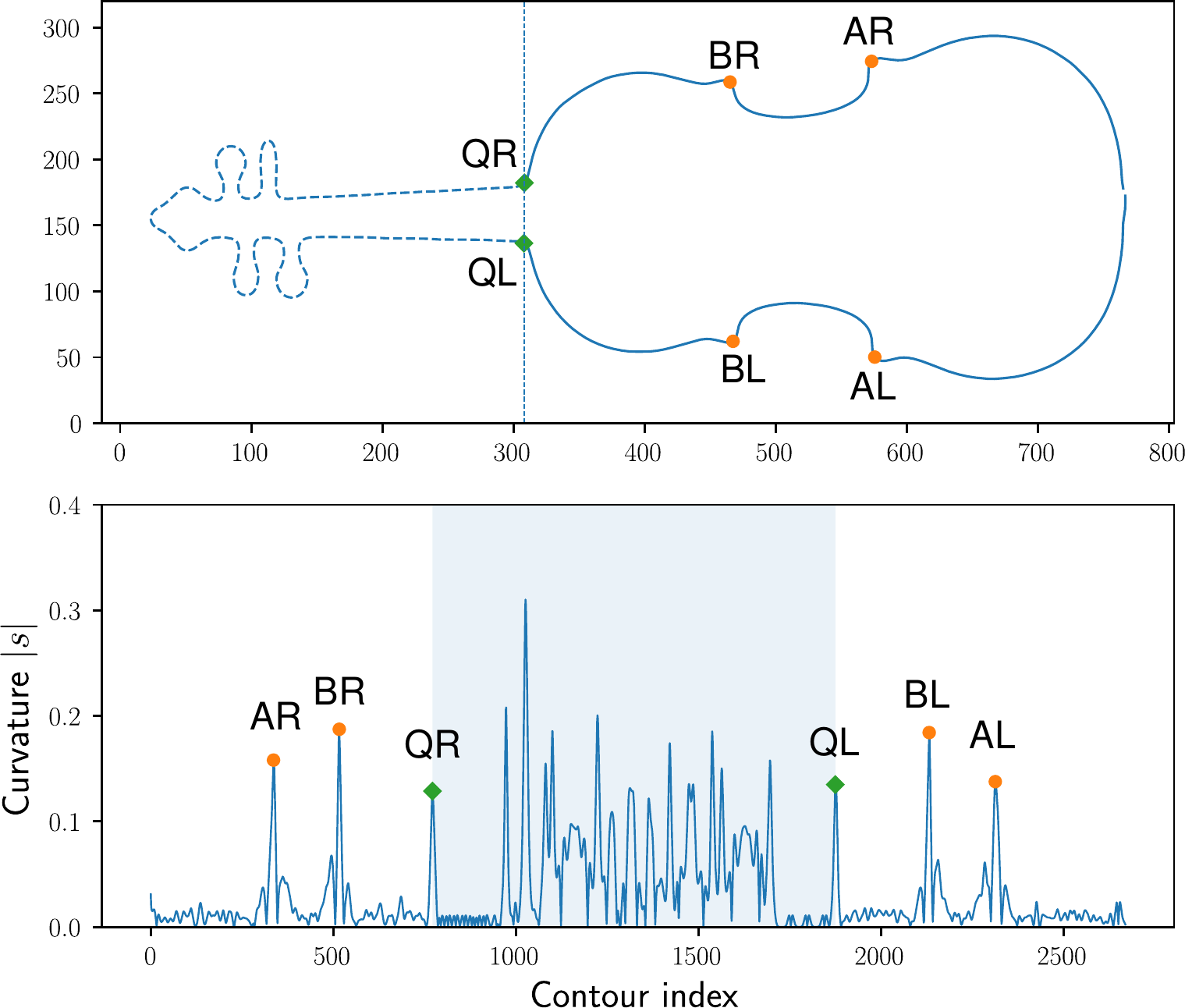}
\end{center}
\caption{(a) Violin contour and (b) its curvature $|s|$ (see Appendix~\ref{sec:appendix_curvature}, Eq.~\ref{eq:curvature}). The control points (AR, BR, AL and BL) are depicted in panel (a) with their associated curvature peaks in (b). Dashed part in the 
contour in (a) represents the fingerboard, pegbox and scroll parts, which are neglected in the study. Shaded area in (b) corresponds to the curvature values associated with the dashed segment in (a). The contour used in (a) belongs to a violin by Franz Zucker~\cite{franzzucker}. }
\label{fig:explaining_detection}
\end{figure}

Once the control points (denoted as AL, AR, BL, and BR in Fig.~\ref{fig:explaining_detection}) 
have been identified via the detection of the first two (AR and BR) and last two (BL and AL) 
peaks in the curve of $|s|$ (see Fig.~\ref{fig:explaining_detection}(b)), the coefficients $a$ to $f$ can then be straightforwardly calculated. However, in order to calculate the average curvature values of the segments in the violin contour, we need to identify two other points, namely QR and QL depicted in Figs.~\ref{fig:measures} and~\ref{fig:explaining_detection}. 
These points mark the position at which the fingerboard projects outside of the violin body. Fortunately,  as we can see in Fig.~\ref{fig:explaining_detection}(b), points QR and QL also have a well defined signature in the curve of $|s|$, i.e. they correspond to the third and fourth peaks along $|s|$. As it is seen in Fig.~\ref{fig:explaining_detection}(b), the contour 
segment between points QR and QL is not considered in the extraction of any measurement because of the difficulty to have the tuning pegs in a standard position and scroll shape variations. 
Thus, in order to not introduce spurious effects, in the analyses that follow, we discard the curvatures associated to the dashed segment in the violin contour in Fig.~\ref{fig:explaining_detection}(a).

\section{Discussion}
\label{sec:discussion}

\subsection{Relationships between violin parts}

\begin{figure}[!t]
\begin{center}
\includegraphics[width=1.0\linewidth]{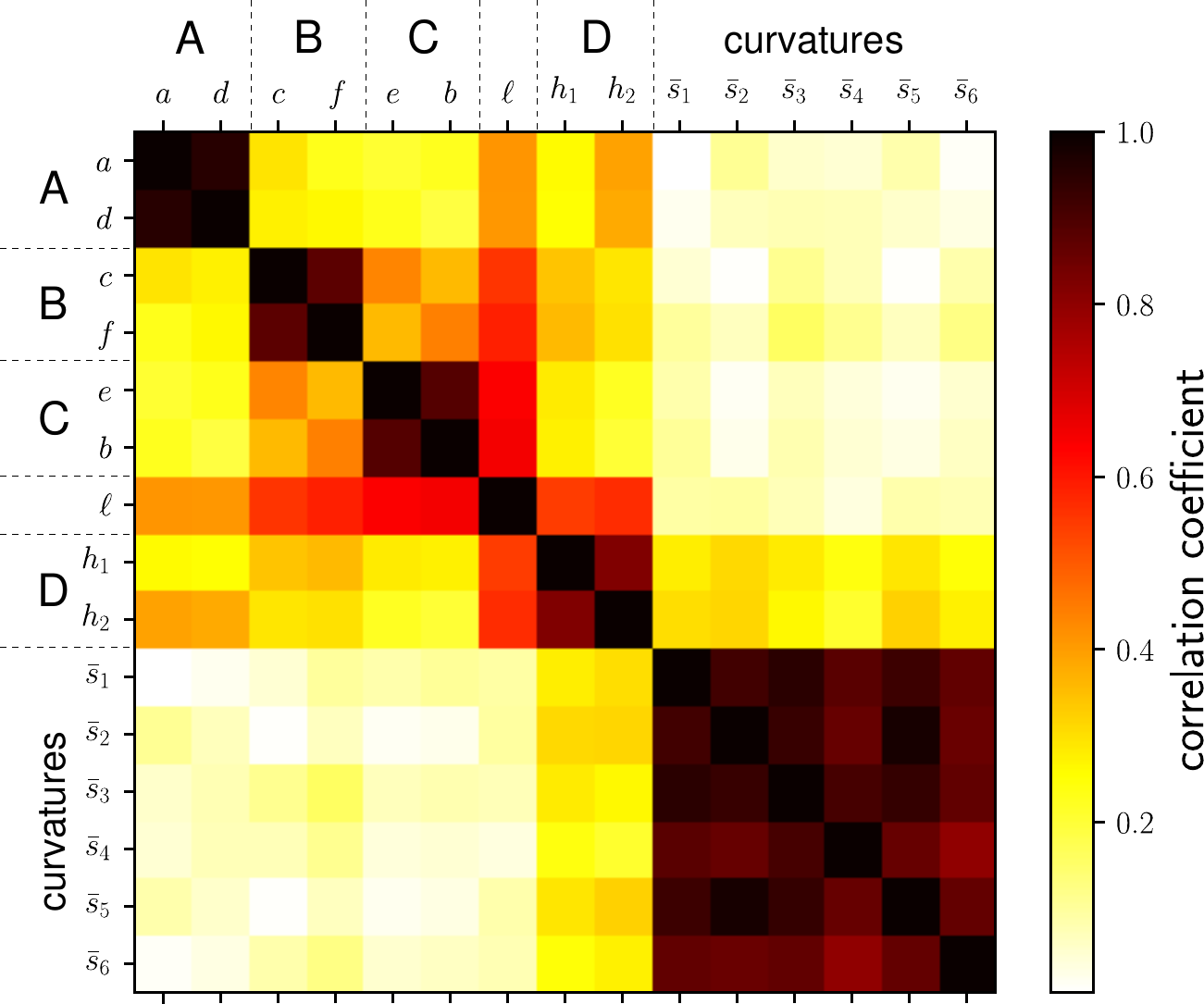}
\end{center}
\caption{Correlation map of the measures defined Fig.~\ref{fig:measures} considering all violins in the database.}
\label{fig:correlations}
\end{figure}

We start our analysis by characterizing the statistical similarity 
between different parts of the violins. More precisely, for each image in the database,  we extract all the measurements depicted in Fig.~\ref{fig:measures} and calculate the 
correlation coefficients between every pair of measurement. 
The resulting correlation map is shown in Fig.~\ref{fig:correlations}. These results reveal interesting patterns of relationships regarding the instruments structural organization. In particular, relevant information can be extracted about the inter-relation between blocks identified in Fig.~\ref{fig:correlations}. For instance,
block $A$, corresponding to the symmetric pair of measurements \emph{a} and \emph{d}, tends to be little correlated with other body features.  On the other hand, measurements in blocks $B$ (measurements \emph{c} and \emph{f}) and $C$ (measurements \emph{b} and \emph{e}) correlate more strongly with the remainder measurements.  This shows that the central ribs of violins are a more independent feature of the instrument, with relative dimensions  conserved among instruments irrespectively of other features.   Still regarding blocks B and C, they present moderate correlation one another, creating a larger cluster with two subgroups.  This suggests that the upper and lower ribs of the instruments are intrinsically interrelated, defining the main outline of the instrument body, with the central ribs acting more as a kind of detail to the overall shape.  However, it is unclear if such phenomena are more a consequence of aesthetic or technical constraints.

Of particular interest is the significantly higher values of correlation between the neck length $\ell$ and blocks $B$ and $C$. In fact, $\ell$ is also strongly related with $D$ (corresponding to measurements $h_1$ and $h_2$) as well.  Such a peculiar characteristic of $\ell$ is reflected in the red line in Fig.~\ref{fig:correlations} extending along most of the blocks defined by the relative dimensions.  As $\ell$ corresponds to the length of the neck, this result indicates that the neck length works together with the other body dimensions for aesthetic, structural or musical reasons.  As such, the neck length assumes  particular importance, defining a reference from which the other dimensions can be derived.

\begin{figure*}[!t]
\begin{center}
\includegraphics[width=0.9\linewidth]{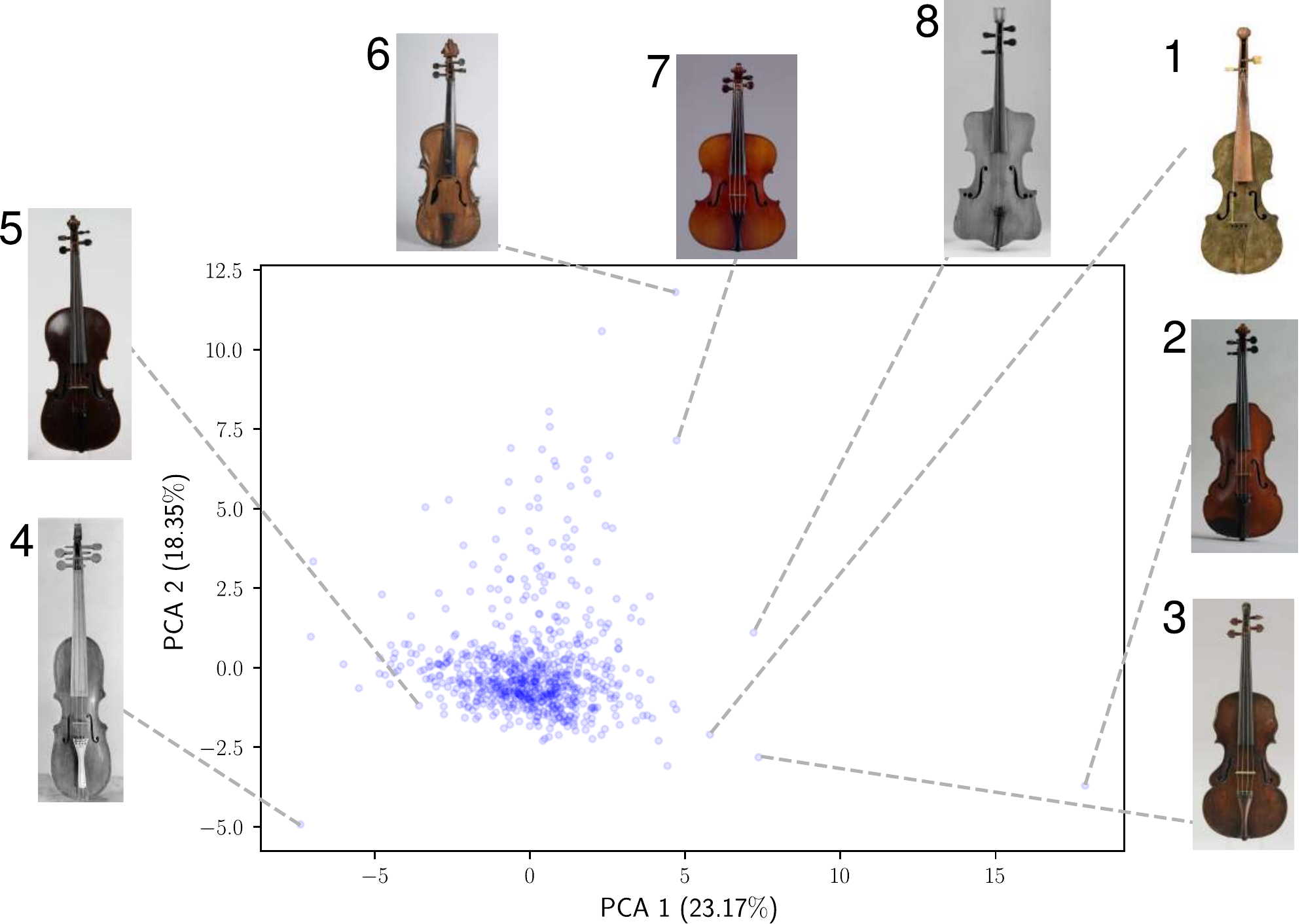}
\end{center}
\caption{PCA projection of the violins considering the whole violin database of this study. Each point in the plot corresponds to an instrument. Violin images reprinted with permission from the Musikinstrumentenmuseums der Universit\"at Leipzig, Cit\'e de la Musique -- Philharmonie de Paris, and Germanisches Nationalmuseum.}
\label{fig:pca_all}
\end{figure*}


Despite the fact that the measurements belonging to group $A$ (i.e. $a$ and $d$) are relatively more independent from the rest, the strong correlations values between the other groups of measurements indicate that the other dimensions preserve an allometric relationship.  There is also a trend towards preserving the aspect ratio (ratios between length and width) of the instruments, as expressed by dependence between $\ell$ with $h_{1}$ and $h_{2}$.  In other words, longer necks require larger separations between the left and right sides of the violin, so that the scale is preserved. 

The curvatures block in Fig.~\ref{fig:correlations} is formed by the mean curvature values of the segments $\gamma_i$ ($i=1,...,6$) marked in Fig.~\ref{fig:measures}. We have that all measurements in this block are intercorrelated in a surprisingly strong fashion. Probably, this intense interrelationship between the curvatures is a consequence of a combination of functional, structural and aesthetics constraints. 

\section{Principal Component Analysis and Classification of violins}
\label{sec:PCA}

\begin{figure*}[!t]
\begin{center}
\includegraphics[width=1.0\linewidth]{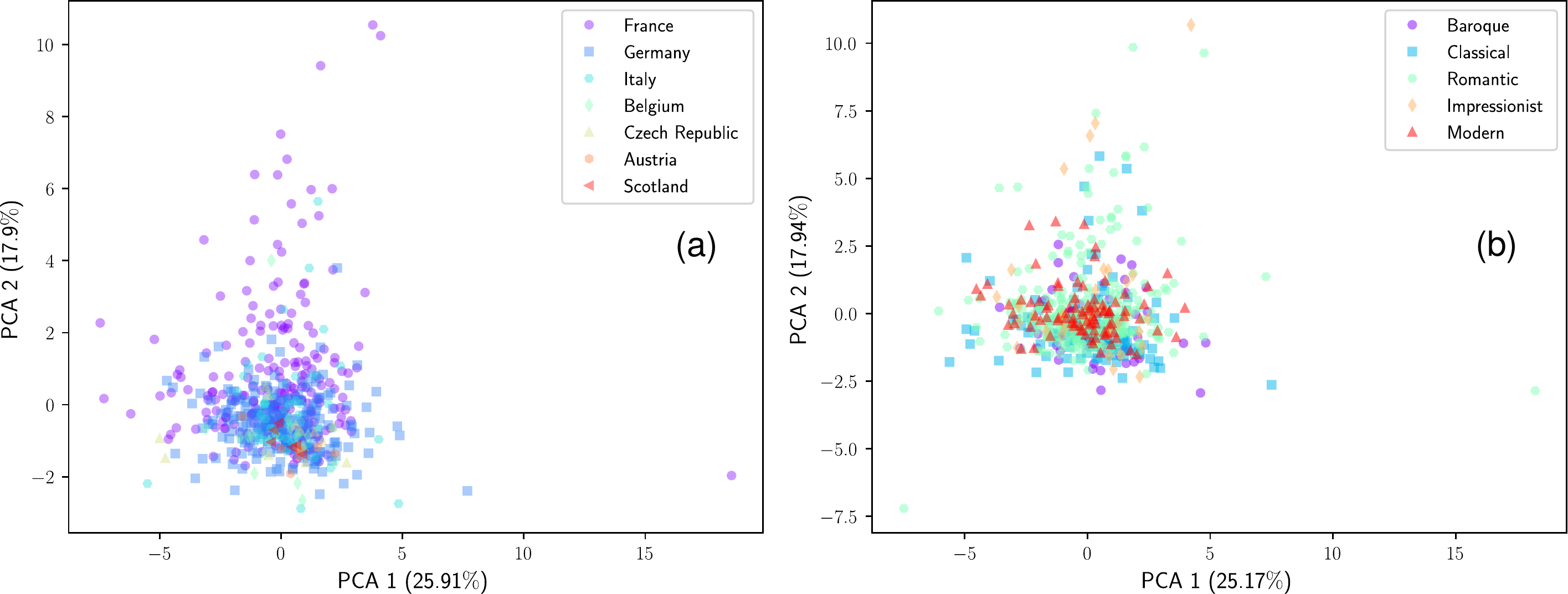}
\end{center}
\caption{PCA projections of the violins considering (a) only violins with information on country of fabrication, and (d) only violins with data of fabrication data. Each point in the plots correspond to an instrument.}
\label{fig:pca_country_period}
\end{figure*}

\begin{figure*}[!t]
\begin{center}
\includegraphics[width=0.8\linewidth]{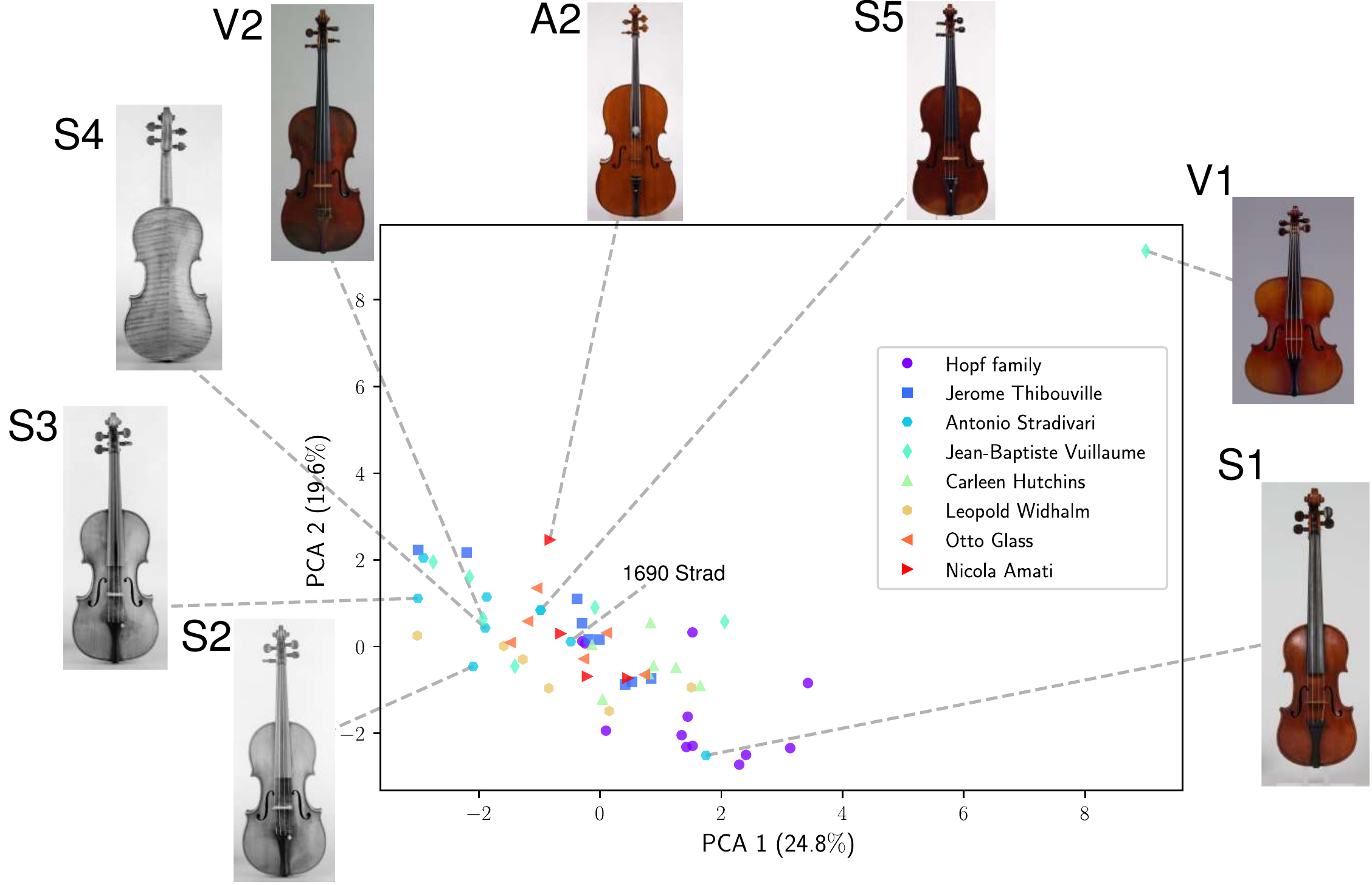}
\end{center}
\caption{PCA projections of the violins considering only violins fabricated by the seven most frequent makers in the database. Each point in the plots correspond to an instrument. Violin images reprinted with permission from the Musikinstrumentenmuseums der Universit\"at Leipzig, Galleria dell'Accademia di Firenze, Cit\'e de la Musique -- Philharmonie de Paris, and Germanisches Nationalmuseum.}
\label{fig:pca_maker}
\end{figure*}

In this section we characterize the distribution of violins 
in the feature space by employing PCA. Importantly, in order to 
have similar ranges of variations, all measurements were normalized prior 
the application of PCA.


Figure~\ref{fig:pca_all} shows the projection of the 728 violins in the space spanned by the two most significant PCA axes considering all measurements. A great part of the violin ensemble gives rise to the large cluster at the origin of the coordinate system. Such an agglomeration suggests a weak geometric variability across the violins belonging to this central group. This is somewhat expected due to the fact that violin making is based on established instruments dimension standards~\cite{johnson1999art}.  Thus, since many instruments are constructed under similar guidelines, it is 
natural to observe such a clustered structure in the PCA space. Nonetheless, 
some outliers can still be observed in Fig.~\ref{fig:pca_all}, particularly at significantly high values of PCA 1 and/or 2. In these
extreme are instruments with unusual relative widths, such as violins 4 and 7; and peculiar shapes as is the case of violins 1 and 3.

In order to try identifying some reason accounting for the dispersion of instruments in Fig.~\ref{fig:pca_all}, in Fig.~\ref{fig:pca_country_period}(a) we visualize the violins according to the their country of fabrication. Only instruments containing country information (572 in total) could be included in this plot, most of which originate from France, Germany, and Italy, in which important makers such as N. Amati, A. Stradivari, J. Vuillaume developed their career.  In addition, reference production and commerce centers were located in such countries, such as Mittenwald in Germany, Mirecourt in France, and Cremona in Italy~\cite{hart1909violin,haweis1979old}.   Interestingly, as we can see in Fig.~\ref{fig:pca_country_period}(a), all the outliers are found to belong to these three countries.  Instruments from other countries tended to overlap violins from the three aforementioned countries, suggesting that European violin design and making could have been heavily influenced by trends originated in France, Germany and Italy.  It is noteworthy observing that the largest dispersion of instruments in the PCA space corresponds to French violins, suggesting greater diversity of design experimentation or co-existence of more than one dominant set of design rules.  

Figure~\ref{fig:pca_country_period}(c) depicts the PCA projection annotated according to the historical periods in classical music. As it can be seen, the vast majority of the violins in the central cluster belongs to Baroque, Classical and Romantic periods.  Interestingly, more recent instruments from the Impressionist and Modern periods resulted mostly concentrated at the core of the central cluster. Figure~\ref{fig:pca_country_period}(c) also implies that violins fabricated after the XIX century have geometrical features closely resembling Baroque and Classical instruments. At first sight, this could be understood as a surprising result, contrasting to the intense experimentation that is characteristic of the modern period.  However, this is more likely a consequence of the fact that a great deal of the music performed at that time, which witnessed wide popularization of music, ultimately originated from the previous periods.

Finally, Fig.~\ref{fig:pca_maker} presents the PCA space with the violins of the 
eight most frequent violin makers in the MIMO database. A good relationship can be observed between the diverse makers and groups of instrument, showing that it is indeed meaningful to study the violin dataset in terms of different types of authorial designs.
Notice, for instance, the consistence of the group formed by the Hopf Family. The Hopf Family is a lineage of violin makers that can be traced back as far as eight generations and that was initiated by Gaspar Hopf (1650-1711)\cite{haweis1979old}.  He is considered one of the founding father of the early luthier tradition in Germany~\cite{haweis1979old}. In the MIMO database, most of the violins belonging to the Hopf group are labeled simply as ``Hopf'', without 
mention to any specific family member, except for two violins attributed to Carl Friedrich Hopf~\cite{hopfviolinsMIMO}. 

\begin{figure*}[!tpb]
\begin{center}
\includegraphics[width=0.7\linewidth]{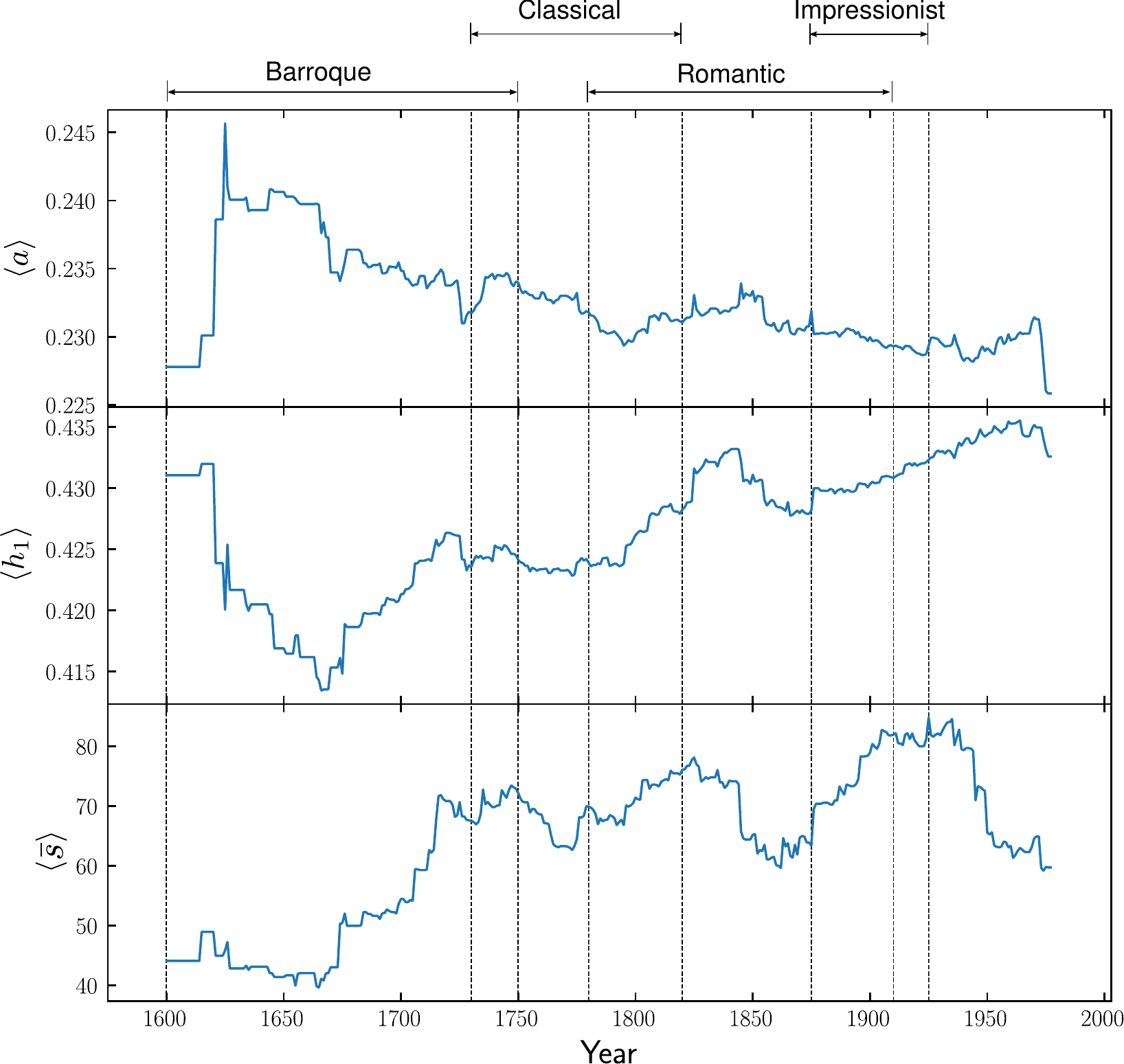}
\end{center}
\caption{Temporal evolution of violin measures (see Fig.~\ref{fig:measures} for measures definition). (a) Average central length $\langle a \rangle$, (b) average distance $\langle h_1 \rangle $ between control points BL and BR, and (c) average curvature $\langle s \rangle$ of violin contours. The averages are calculated by considering time-windows with length $\Delta t = 20$ years (see text for details) and sliding step $\delta t = 1$ year.}
\label{fig:time_series}
\end{figure*}

It is worth dedicating some attention to the Stradivari's violins, which occupy a central 
position in the PCA space, strongly overlapping other groups. Additional insights can be gained by observing the results of Fig.~\ref{fig:pca_maker} in the light of important landmarks along the history of these famous instruments. For instance, it is known that between the years 1656--1684, Stradivari was an apprentice to Nicola 
Amati's workshop in Cremona, Italy~\cite{StradivariLiveNWork}. In this period (oftentimes referred as Amatis\'e period~\cite{StradivariLiveNWork}), Stradivari developed violins following ``Amati's grand pattern''~\cite{StradivariLiveNWork}. Remarkably, by inspecting Fig.~\ref{fig:pca_maker} one finds that the Stradivari instrument closest to an Amati is a violin dated from 1690 (see Fig.~\ref{fig:pca_maker}). The other Strads in the PCA space belong to periods later than 1690, the year in which
Stradivari sought to depart from Amati's influence, and began the production of the so-called Long Strads~\cite{StradivariLiveNWork}. 

The Long Strads have significantly larger patterns than the traditional styles of the Cremona school, being so designed in order to meet other demands of power and 
tonality~\cite{StradivariLiveNWork,hutchins1983history}. Eventually, Stradivari started producing small violins in addition to other sizes. One example is violin S1 in the lower left part of the cluster in Fig.~\ref{fig:pca_maker}. This violin is significantly smaller than the other instruments by Stradivari (compare it, for instance, with S5): S1 has a total length of 53.5 cm, while the other Strads have lengths varying between 
58.8-59.7 cm. This reduction in size was followed by changes in the proportional measurements as indicated in Fig.~\ref{fig:measures}. For instance, S1 has $a = 0.26$, whereas the remaining Strads in Fig.~\ref{fig:pca_maker} have measure $a$ in the range $a \in [0.21,0.22]$.

A further consideration is worth regarding the relationship between 
Stradivari's and J. Vuillamme's instruments. Observe that most part of violins by these two makers resulted nearby one another in the PCA space of Fig.~\ref{fig:pca_maker}.  This result is particularly interesting as it confirms in a quantitative way the well-documented influence of Stradivari on Vuillaume~\cite{haweis1979old}. In fact, not only the latter is known to have been a prolific Strad copyist, but he also produced remarkable replicas of other violins belonging to the Cremona school. Some of these copies were hardly distinguishable from 
the originals, even by highly skillful violinists such as Niccolo Paganini, 
who once was unable to distinguish between a violin from the Cremonese school and its counterfeit crafted by Vuillaume~\cite{haweis1979old}. The proximity of the points
in the PCA thus agrees with the historical narratives of the mentioned 
luthiers.

Another relevant point in the analysis of violin shapes concerns the importance of each 
measurement in predicting characteristics such as country of fabrication and maker. This 
task can be addressed by using Random Forest Classifiers~\cite{breiman2001random,james2013introduction}. First, we apply this technique 
to the violin data used in the PCA projection visualized in terms of countries (Fig.~\ref{fig:pca_country_period}(a)). 
By randomly splitting 
the data into 70$\%$ for training and 30$\%$ for testing, we are able to achieve 57$\%$ of accuracy 
in predicting the country of fabrication, whereas the random case returns $16\%$ of correct assignment. 
By repeating the procedure for the data of Fig.~\ref{fig:pca_maker}, we get $44\%$ of accuracy 
in the prediction of violin makers, over $12\%$ in the random case. 







\section{Temporal analysis and Contour Morphing by Thin Plate Splines}

\label{sec:temporal_analysis_morphing}
\begin{figure*}[!tpb]
\begin{center}
\includegraphics[width=0.7\linewidth]{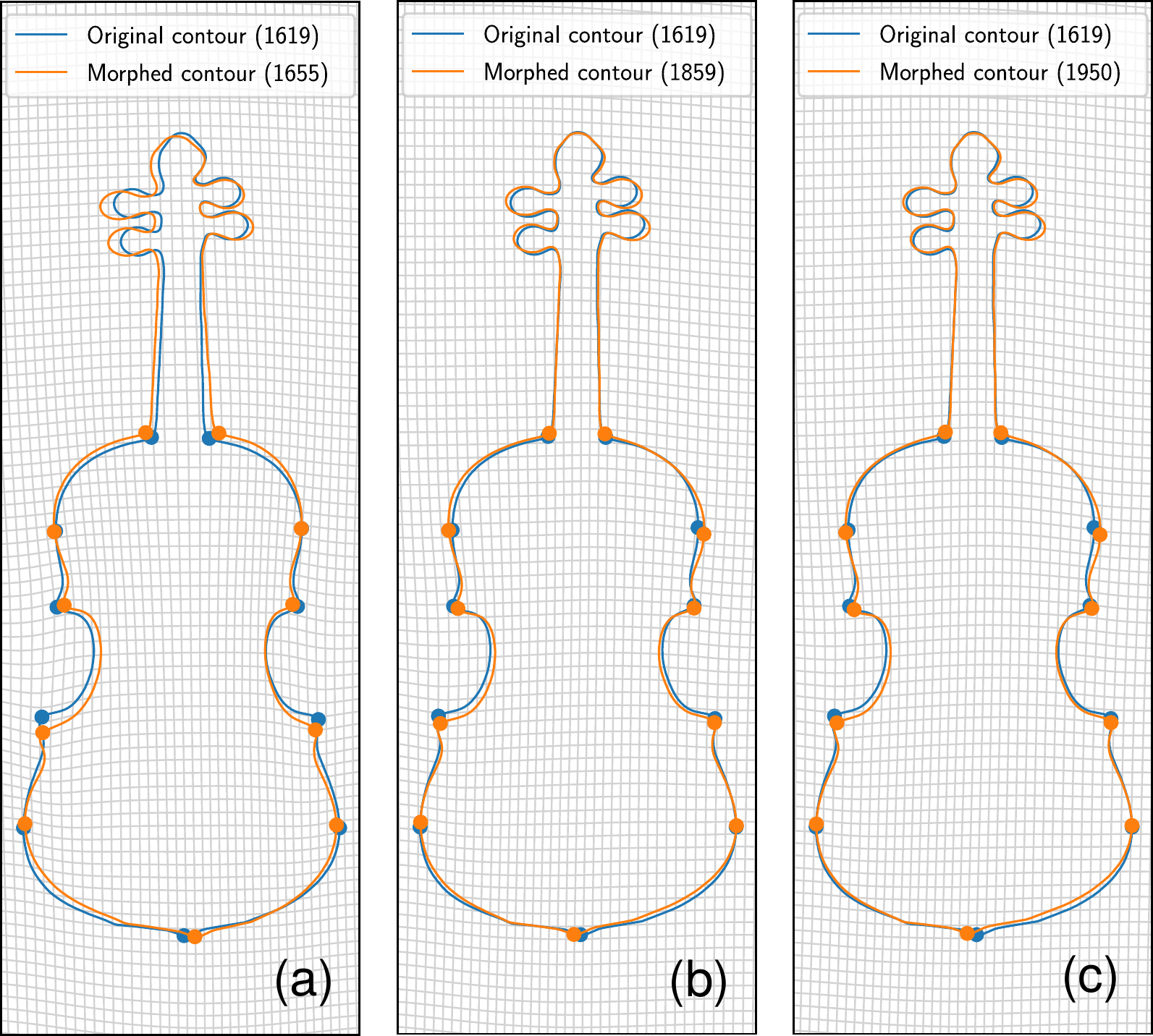}
\end{center}
\caption{Morphing of a violin fabricated in the year of 1619 to the average shapes of the time windows
centered at (a) 1655, (b) 1859, and (c) 1950.}
\label{fig:morphing}
\end{figure*}

The PCA projection in Fig.~\ref{fig:pca_country_period}(b) reveals 
the statistical variability in the violin design across 
epochs in classical music history. However, 
although this result quantifies how different instruments 
fabricated in distinct periods can be, it does not 
tell us exactly how the violin body changed over time. 
In this section we analyze the evolution 
of the violin shape by employing two approaches, namely (i) 
the visualization of the time series associated to the measures 
defined in Sec.~\ref{sec:materials_and_methods}, and (ii) the study of the morphing of 
contours by using the \emph{thin-plate splines} method~\cite{costa2001shape}. 

The oldest violin in the MIMO database goes back to 1575, whereas the newest one was fabricated 
in 2003. Our goal here is to investigate changes in the average length- and curvature-based measurements, corresponding to a ``characteristic 
violin'', along the above time spam.  
To this end, we make use of a sliding time window technique.  The adoption of windowing is necessary in order to provide a prototypic (average) characteristic for a time period and also to reduce high frequency noise in the time series.  First, we set a time window of length $\Delta t = 20$ years inside 
which the quantities of the violins belonging to the covered period are averaged. By subsequently 
sliding this time window by $\delta t = 1$ year, we generate a time series.   More specifically, the first element in the time series associated to, for instance, quantity $a$ is calculated by averaging the corresponding values of the violins that were fabricated between the years $y_1^{(1)} = 1575$ and $y_2^{(1)} = 1575 + \Delta t = 1625$. Next, the second element of the time series is obtained by averaging $a$ over the violins belonging to the period
between years $y_1^{(2)} = 1575 + \delta t = 1576$ and $y_2^{(2)} = y_1^{(2)} + 1626$. This process is repeated for all possible intervals. Furthermore, each time window is labeled with its midpoint year in order to provide
a time reference for the averaged values. Henceforth, we denote by $\langle \cdot \rangle$ averages calculated
over violins that fall in the same time-window.

Figure~\ref{fig:time_series} shows the evolution of some average violin measurements obtained
according to the methodology discussed above. As it can be seen, quantities 
$\langle a \rangle $ and $\langle h_1\rangle $ exhibit slight
decreasing and increasing trends, respectively. The average curvature 
$\langle \bar{s} \rangle$, on the other hand, shows oscillatory 
behavior, but with small amplitudes of variation. This supports 
the conclusions derived with PCA projection in Fig.~\ref{fig:pca_country_period}(b)  that violin design remained mostly stable 
across time, without adhering to particular trends over the different periods in classical music 
history. 

Another way to visualize changes in violin shape is by 
means of thin plate splines~\cite{costa2001shape}. Such a method was also employed in~\cite{chitwood2014imitation}
so as to depict the morphological deformations required to transform 
a representative violin outline of a given luthier into a violin shape 
of another violin maker. Here, we consider the thin plate spline 
technique in order to construct a morphing transformation from 
a given violin shape to a target contour that represents 
the average shape of a certain time period. In other words, 
we seek for the necessary morphological deformation to 
transform a given violin into the average shape of a certain epoch. 
Our methodology works as follows: first, a violin 
shape from the MIMO database as the reference contour is selected. Second,
for each sliding time window, the average control contour points are calculated 
and a time evolving set of thin plate splines transformations is obtained. 

In Fig.~\ref{fig:morphing} we show the morphing via thin plate splines of a violin 
fabricated in 1619 having as targets three periods, namely the time windows centered 
at 1655, 1859 and 1950. As it can be seen, only a small difference is observed 
in the position of the control points. The most prominent displacement 
of the grid corresponds to the year of 1655, where we observe a slight dilation 
of the two most upper control points (see Fig.~\ref{fig:morphing}(a)). In the 
Supplemental Material, we provide a video with the animated version of Fig.~\ref{fig:morphing}, which
shows the complete evolution of the morphing over the time span covered by the 
MIMO database. At first sight, this result seems to contradict the findings in~\cite{chitwood2014imitation}, where a 
negative correlation between time and Linear Discriminant Analysis (LDA) coefficients associated to violin shapes was reported. However, in the present work, correlation with time was also observed in the violin time series, but with moderate amplitudes of variation, which yield small grid deformations in the time averaged contours.

\section{Conclusion}
\label{sec:conclusions}

The primary goal of this paper was to statistically characterize and study
the violins stored in the MIMO dataset. The interest in this platform resides in the fact that it contains the largest freely available online collection of instruments~\cite{MIMO}. Aiming at capturing relevant geometrical 
features of the violin body, we initiated our study by defining 
measurements that were divided into two groups: length-based and curvature-based.

The analysis of the extracted features by means of PCA revealed interesting relations among 
violin makers. In particular, it was confirmed~\cite{chitwood2014imitation} that famous violin makers tend to cluster 
with their respective copyists, such as in the case of Stradivari and Vuillaume. Another interesting
result concerns the proximity in the PCA space between Stradivari and Amati instruments: it was shown that the
closest Strad to an Amati corresponds to a violin crafted in 1690,  which is precisely 
the year that Stradivari changed his design from the ``Amati's grand pattern'' to the production 
of the Long Strads. Moreover, influence between masters and apprentices was reflected in low statistical variability among violins originated from the Hopf family, the
only lineage of luthiers represented in the MIMO database. Such a relationship between 
luthiers, copyists and family lineages was also reported for the database studied in~\cite{chitwood2014imitation}.

By using thin plate spline analysis, we found that the average violin
shape changed little along time -- a result
corroborated by the lack of a clustered structure in the PCA as visualized in terms of the period of fabrication. The contour morphing showed, with few exceptions (see the video in the supplemental material), 
that small deformations in the grids are enough to transform standard shaped violins to the average outlines 
respective to different epochs.

All in all, besides providing a quantitative characterization 
of the violin dataset of MIMO, our analysis also complements previous studies of violin shape contrasted with historical information as reported in~\cite{chitwood2014imitation}. This paves the way to further research encouraging, in the process, more  museums to exhibit their collection through online platforms. 




\section*{Acknowledgements}

We gratefully acknowledge the MIMO initiative 
for providing the images that were used in this 
study. We also greatly acknowledge Daniel Wetterskog 
(Director of the Swedish Museum of Performing Arts), Cecilie Hollberg (Director 
of the Galleria dell'Accademia di Firenze), Josef Focht (Director of the Musikinstrumentenmuseums der Universit\"at Leipzig), Philippe Provensal (Head of Press Department, Cit\'e de la Musique -- Philharmonie de Paris), and Bianca Slowik (Germanisches Nationalmuseum), 
for kindly authorizing the reprint of the violin images used in Fig.~\ref{fig:contour_example},~\ref{fig:pca_all} and~\ref{fig:pca_maker}. TP thanks Filipi Nascimento, Cesar Comin, and Lucas Assirati for useful discussions. 
TP is further grateful to FAPESP for sponsorship, FAR acknowledges the Leverhulme Trust, CNPq (Grant No. 305940/2010-4) and FAPESP (Grants
No. 2016/25682-5 and grants 2013/07375-0), and LdaFC thanks CNPq (grant No. 307333/2013-2) and NAP-PRP-USP for sponsorship. 

\appendix

\begin{figure}[!tpb]
\begin{center}
\includegraphics[width=1.0\linewidth]{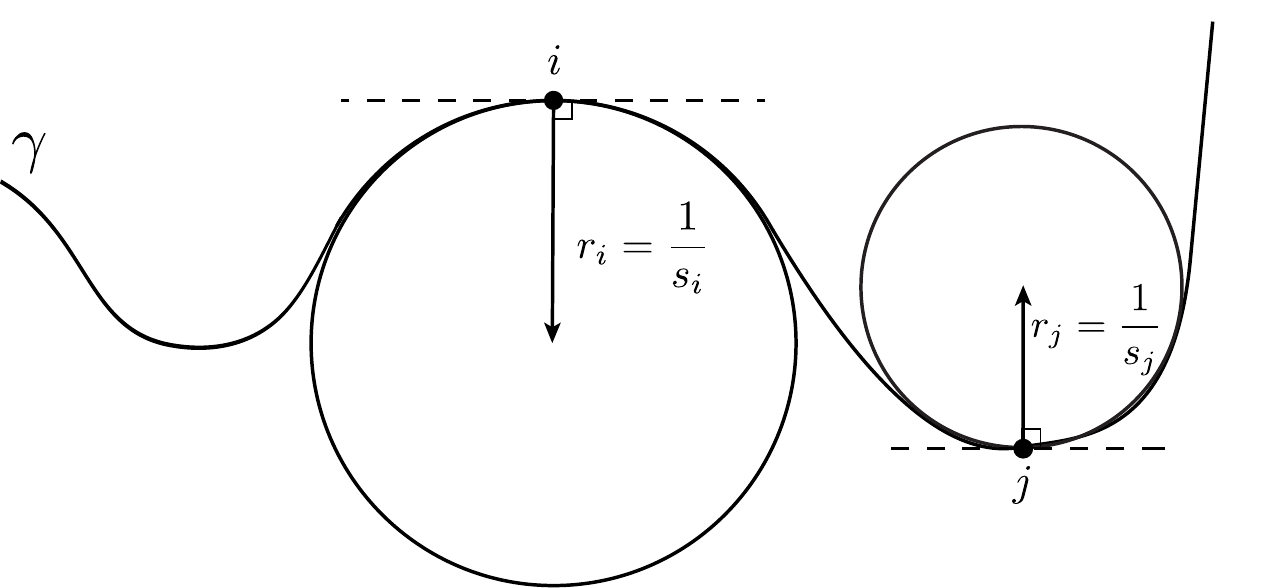}
\end{center}
\caption{Illustration of the concept of curvature (Eq.~\ref{eq:curvature}) of an arbitrary planar curve $\gamma$. The curvature value at a point $i$ in $\gamma$ is given by the inverse of the radius $r_i$ of the osculating circle at $i$, i.e $s_i = 1/r_i$. In the 
above example, point $j$ has higher curvature than $i$.}
\label{fig:curvature}
\end{figure}

\begin{figure*}[!t]
\begin{center}
\includegraphics[width=0.8\linewidth]{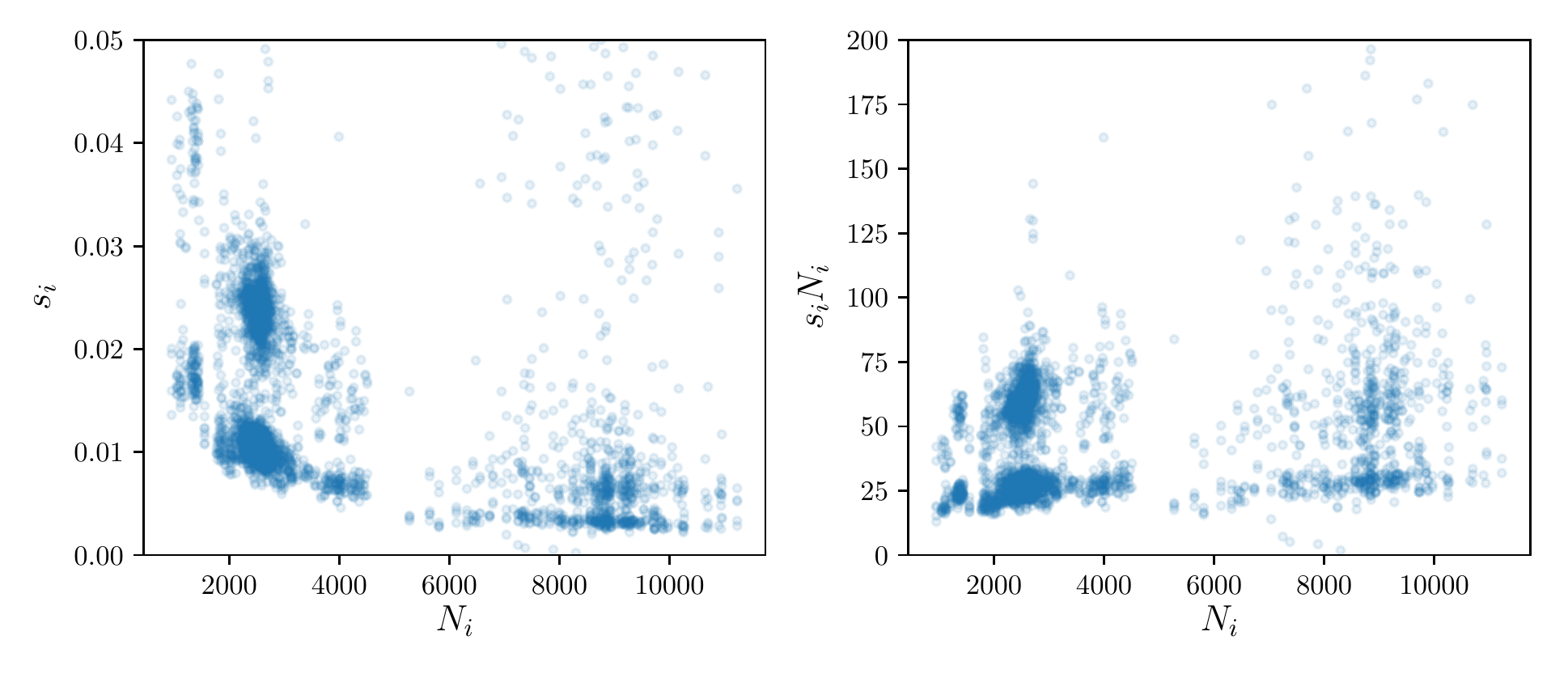}
\end{center}
\caption{Normalization of curvature-based measurements. (a) Scatter plot showing 
$\bar{s}_i$ as function of the number of points $N_i$ in a contour segment $\gamma_i$. (b) Dependence of the normalized averaged curvature (Eq.~\ref{eq:normalization_curvature}) on $N_i$.}\label{fig:normalization}
\end{figure*}

\section{Curvature Definition}
\label{sec:appendix_curvature}


Given a planar, regular, parametric curve expressed in terms of Cartesian coordinates $\gamma(t) = (x(t),y(t))$, its curvature is defined~\cite{costa2001shape} as
\begin{equation}
s=\frac{\dot{x}\ddot{y}-\dot{y}\ddot{x}}{(\dot{x}^{2}+\dot{y}^{2})^{\frac{3}{2}}}, 
\label{eq:curvature}
\end{equation}
where the dots denote the derivatives with respect to parameter $t$. Informally speaking, the curvature $s(t)$ quantifies the rate at which the orientation of the tangent vector changes along $\gamma$. This can be intuitively understood as an indicator of the local geometric nature of an arbitrary curve. For instance, regions with constant $s=0$ correspond to straight line segments in the original curve. On the other hand, if a portion of $\gamma$ exhibits non-null constant $s$, then this region corresponds to a perfect circle. In fact, the value of $s$ at a certain point is given by the inverse of the radius of the osculating circle to that point~\cite{costa2001shape}. This is exemplified in Fig.~\ref{fig:curvature}.

\section{Measurements Normalization}
\label{sec:appendix_normalization}

The measurements proposed in the main text for the violin characterization are divided into two sets: measurements based on the length and width of violin parts ($a$, $b$, ..., $f$; $h_1$ and $h_2$), and measurements derived from the curvature of the segments $\gamma_i$ ($\bar{s}_1$,..., $\bar{s}_6$). However, as they are defined, these quantities cannot be employed to compare different violins because they clearly depend on the sizes of the original images from which they were obtained. More specifically, length-based measures are directly proportional to the width of the images, while 
the average curvatures scale with the contour size as depicted in Fig.~\ref{fig:normalization}(a).

In the case of the length-based measures, this problem is circumvented by normalizing the measurement values by the total violin's length $L$, i.e.  
\begin{equation}
\tilde{a} = \frac{a}{L},
\label{eq:normalization_lengthbased}
\end{equation}
and analogously for $\tilde{b}$, ..., $\tilde{f}$,$\tilde{h}_1$ and $\tilde{h}_2$.

Unfortunately, the normalization in Eq.~\ref{eq:normalization_lengthbased} cannot
be extended to curvature-based measurements. It turns out that the coefficients 
$\bar{s}_i$ are inversely proportional to the number of points $N_i$ in segment $\gamma_i$ (see Fig.~\ref{fig:normalization}(a)). To correct this effect and obtain non-dimensional measures we define  
\begin{equation}
\tilde{ \bar{s}}_i  = \bar{s}_i N_i, 
\label{eq:normalization_curvature}
\end{equation}
As shown in Fig.~\ref{fig:normalization}(b), by doing so, we eliminate the correlation between $\bar{s}_i$ and the length of the respective contour segments. For simplicity's sake, in the main text we omit the upper scripts ``$\sim$'' from the
variables notation.

\bibliography{bibliography}

\end{document}